# Spin-Wave-Assisted Thermal Reversal of Epitaxial Perpendicular Magnetic Nanodots

S. Rohart,[1,][*] P. Campiglio,[2] V. Repain,[2,][†] Y. Nahas,[2] C. Chacon,[2] Y. Girard,[2] J. Lagoute,[2] A. Thiaville,[1] and S. Rousset[2]

[1]*Laboratoire de Physique des Solides, CNRS, Université Paris Sud, UMR 8502, Bâtiment 510, F-91405 Orsay Cedex, France*
[2]*Laboratoire Matériaux et Phénomènes Quantiques, Université Paris Diderot-Paris 7, UMR CNRS 7162,
10 rue Alice Domon et Léonie Duquet 75205 Paris Cedex 13, France*

The magnetic susceptibility of self-organized two-dimensional Co nanodots on Au(111) has been measured as a function of their size in the 2–7 nm diameter range. We show that the activation energy for the thermal reversal displays a power law behavior with the dot volume. Atomic scale simulations based on the Heisenberg Hamiltonian show that this behavior is due to a deviation from the macrospin model for dot size as small as 3 nm in diameter. This discrepancy is attributed to finite temperature effects through the thermal excitation of spin-wave modes inside the particles.

The thermal stability of nanoparticles is a forefront challenge for the realization of ultrahigh density recording media. To overcome the so-called superparamagnetic limit [1], intense effort has been made for finding new high magnetic anisotropy materials. However, it appears that nanoparticles can display rather different properties from the bulk and that the achievement of magnetic nanoparticles with long-term stability at room temperature is still pending. This objective requires one to fully understand the physics of the magnetization thermal reversal in nanostructures. In classical spin systems, it is often described by the Néel-Brown theory of superparamagnetism, in the framework of the macrospin approximation [2,3]. The reversal is therefore supposed to be fully coherent; i.e., all the spins are collinear during the magnetization switching and the associated activation energy is the particle magnetic anisotropy energy (MAE). Recently, measurements on single nanoparticles have allowed a direct comparison between experimental data and this model [4,5]. Surprisingly, up to now, its range of validity has only been discussed on the basis of static micromagnetism [6–8], which shows that for a circular nanodot with perpendicular magnetization, the magnetization reversal is coherent for diameters below several exchange lengths. However, the magnetization reversal is intrinsically a dynamical phenomenon [3] and spin-wave excitations can play an important role in this process [9]. Finally, the effect of a finite temperature on the excited magnetic modes and its consequences on the thermal reversal has never been investigated at a nanometer scale.

In this Letter, we investigate the limitation of the macrospin model at finite temperature and provide evidence of the crucial role of thermal spin-wave excitations on the magnetization reversal dynamics. By comparison between experiments and atomic scale simulations, we show that the magnetization thermal reversal of self-organized two-dimensional Co particles deviates significantly from the macrospin expectation above a 3 nm diameter size, i.e., below the exchange length.

Co/Au(111) is a self-organized system that allows an unambiguous interpretation of experiments averaged over a macroscopic surface [10,11]. Indeed, below a Co coverage around 1.6 monolayers (ML) that corresponds to the coalescence to a continuous thin film with double layer growth [12], this system provides an assembly of highly homogeneous bilayer islands of adjustable size in the 2–7 nm diameter range. The nanodots are magnetically independent from a coupling point of view and display an out-of-plane magnetization. Our experiments have been performed using a Au(111) oriented single crystal sample, displaying a mirrorlike surface and 300 nm wide terraces. In the same UHV setup ($P = 5 \times 10^{-11}$ mbar), we perform the Co deposition at room temperature, STM images and polar magneto-optical Kerr effect measurements from 20 to 300 K. The magnetic susceptibility has been measured with a 1 Hz ac external field (10 mT rms amplitude) perpendicular to the sample surface [11,13]. The real and imaginary parts of the dynamical susceptibility are measured with a lock-in amplifier through a crossed polarizer configuration. The experiments have been repeated several times in different conditions with a very good reproducibility. Figure 1(a) shows the susceptibility measurements for an average Co dot size in the 2–7 nm diameter range (the mean diameter being deduced from the experimental area distribution, considering circular islands). The continuous lines show the best fits using a model based on the linear response theory of a two level system including an activation energy $E_a$ and an attempt frequency $\nu_0$ of 40 GHz, combined with a Gaussian distribution of $E_a$ among the particles assembly. Figure 1(b) shows the variation of the average activation energy as function of $N_{\text{at.}}$ the mean number of atoms (the $E_a$ distributions were deduced from the fit leading to a standard deviation $\sigma/E_a \approx 30\%$). In the inset, STM images for five different measurements show the evolution of the dot morphology with the Co coverage. It is important to note that the highest coverage studied (0.7 ML) is well below the coalescence regime and that the dipolar coupling between the dots is negligible

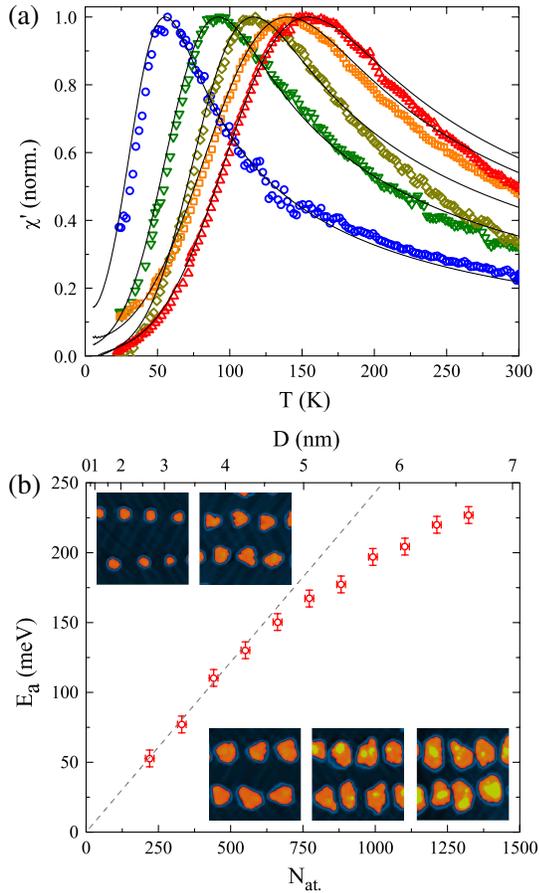

FIG. 1 (color online). (a) Real part of the experimental dynamical magnetic susceptibility $\chi'$ as function of temperature for different mean Co dot sizes (3.3, 4.3, 5.1, 5.7, and 6.3 nm from left to right). In continuous line, fits based on the model described in the text. (b) Average activation energy $E_a$ versus the average atom number per Co dot $N_{at.}$. In inset, $27 \times 27$ nm$^2$ STM images corresponding to an average dot size of 2.7, 4.3, 5.4, 6.0, and 6.6 nm. The nanodots have bilayer height. The bright spots on the larger dots correspond to a third layer nucleation.

($\mu_0 H \approx 40$ mT) as compared to the MAE in this system ($\mu_0 H_K \approx 5$ T), due to the perpendicular magnetization and the bilayer height of the Co islands. Moreover, it is worth noting that few Co atoms can be observed on the top of some islands above 0.3 ML (brighter dots in inset images). Taking into account the tip convolution, that is certainly high for such small lateral structures, we have estimated that they represent only 5% of the total coverage for 0.7 ML, and therefore we have considered their contribution to the magnetization reversal negligible. As it is clearly evidenced by Fig. 1(b), the activation energy does not vary linearly with $N_{at.}$. Whereas it increases linearly for small sizes (almost 0.3 meV/at.), it is deviating for larger sizes, which have smaller activation energies than expected. As we show in the following, this behavior can be perfectly reproduced with an out-of-plane homogeneous anisotropy, considering the important role of thermally excited magnetic modes on the reversal process.

The magnetic nanodots are described at the atomic scale using a Heisenberg Hamiltonian including all the magnetic energies (exchange, anisotropy, and dipole-dipole interaction) and the magnetization dynamics is derived by integrating the Landau-Lifshitz-Gilbert equation [14]. Thermal effects have been included through a random field with a white spectrum [3]. The parameters are magnetic moment $\mu_{at.} = 2.1\mu_B$/at. [15], exchange constant $J = 29$ meV/bond [15], and damping factor $\alpha = 0.2$ [16]. The magnetic anisotropy easy axis is along the dot axis, and its value has been taken homogeneous over the whole particle (0.4 meV/at.). Taking into account the shape anisotropy [14], the activation energy for a coherent magnetization reversal is about 0.3 meV/at. as observed in the low diameter range in the present experimental study. In order to get a reasonable calculation time, we have considered single layer nanodots with a circular shape, the atoms being on a hexagonal lattice. Ten particles have been studied from 1 to 823 atoms, i.e., from 0.2 to 7.6 nm diameter size. In this size range, i.e., below several exchange lengths ($\Lambda \approx 3.3$ nm [14]) and domain wall widths ($\delta \approx 3$ nm [14]), we have checked that the field induced magnetization reversal is coherent at $T = 0$ K and the measured switching field exactly corresponds to the anisotropy field $H_K$ when the field is applied along the easy axis.

The magnetization fluctuations have been simulated in a 100 ns time window and in zero applied field [Fig. 2(a)]. At finite temperature, a reduction of the dot magnetization modulus is observed, indicating that the nanoparticle is in an excited state. Above a threshold temperature, that can be seen as a blocking temperature, superparamagnetic fluctuations are resolved giving rise to a telegraph noise for the

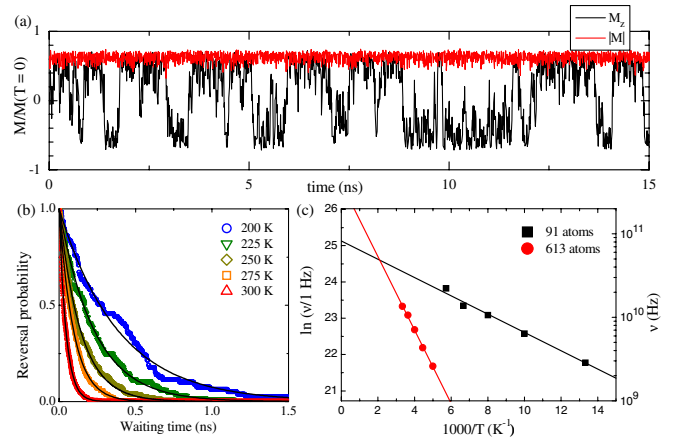

FIG. 2 (color online). (a) Time dependence of the magnetization (normalized to the magnetization at zero temperature) in a 613-atom nanodot at 200 K obtained from the simulations. The gray (red) curve represents the magnetization modulus and the black curve represents the perpendicular component, showing a telegraph noise. (b) Waiting time statistics at different temperatures for the same particle. The lines correspond to exponential decay fits. (c) Arrhenius plot of the switching rate for 91- and 613-atom particles.

perpendicular magnetization component ($M_z$). Between two magnetization reversal events, the magnetization direction is nearly constant. Moreover, during the magnetization change, the amplitude of the total magnetization is constant, indicating an apparent coherent magnetization reversal. A statistical analysis has been performed in order to determine the magnetization reversal probability. This waiting time analysis [4,5] shows in each simulation a single exponential law [Fig. 2(b)] as expected in the Néel-Brown theory [2,3], which allows us to determine the mean switching rate $\nu$. For each particle, a temperature range is determined in order to observe 10–100 reversal events in the 100 ns time window. This allows us to study the variation of $\nu$ versus temperature. All the particles display an Arrhenius law $\nu = \nu_0 \exp(-E_a/kT)$ as shown in Fig. 2(c).

Particular attention has been given to the single atom particle since this case has an exact analytical solution [3]. A good agreement was found with Brown's model, and in that particular case, the activation energy (0.39 meV) corresponds to the MAE (0.4 meV), typical of a macrospin magnetization reversal. For particles up to 100 atoms, the activation energy still coincides with the MAE (Fig. 3). For larger particles a significant deviation is found and the variation of $E_a$ follows a power law ($E_a \sim N_{at.}^{0.5}$).

This result is in a qualitative good agreement with the experiments, showing the two regimes previously described. However, due to the monolayer height in the simulations, the number of atoms, above which deviations to the macrospin model are observed, is different from the experiments. As we show in the following the key parameter is rather the diameter. This provides a much better agreement between experiments (deviation above $\approx 4.7$ nm) and simulations (deviation above $\approx 3.5$ nm), the small difference being attributed to the difference of exchange constant between monolayer and bilayer films [14]. An interpretation in terms of enhanced MAE at the edge atoms [13] would lead in our case to an edge anisotropy of 0.75 meV/at., very close to the values obtained on monolayer height Co dots on Pt(111) [13]. However, this interpretation is not correct here as the MAE in the calculation is identical on every atom. This result rather shows a deviation to the macrospin model.

The deviation to the macrospin is further evidenced looking at $\nu_0$ (inset of Fig. 3). For our parameters, $\nu_0$ is expected to be 40 GHz in the Brown's model [3]. The one atom simulation reproduces perfectly this value. Up to 100 atoms, $\nu_0$ is slightly larger (70–100 GHz). Above 100 atoms, a significant increase is observed with frequencies of several hundreds of GHz. This increase coincides with the deviation of $E_a$ from the macrospin model. Note that the latter result cannot be explained by a reduction of the mean magnetic anisotropy, which would induce a decrease of $\nu_0$. This forbids constructing an effective macrospin model for this system.

As shown in Fig. 2(a), a reduction of the magnetization amplitude is induced by the thermal fluctuations, and is mainly due to incoherent fluctuations from one spin to another. Collective excitations are also present, the most important one being the global magnetization precession, at the origin of the magnetization reversal. In order to observe these excitations, simulations have been performed for a large time (250 ns) and at reduced temperatures where no superparamagnetic fluctuations are observed. A Fourier transform analysis of the magnetization components for each spin in the particle has been performed in order to determine the eigenmodes that are excited. For each mode, the power spectral density (PSD) was mapped. We have considered two nanodots with different sizes and identical reduced temperature as compared to the MAE. For the smaller one [in the size range where $E_a$ varies linearly with $N_{at.}$, Fig. 4(a)], containing 61 atoms (i.e., 2 nm diameter), the precession is homogeneous and the frequency (129 GHz) is in agreement with the macrospin expectation $\gamma_0 H_K/2\pi = 130$ GHz (with $\gamma_0$ the gyromagnetic factor). For the larger one [in the size range where $E_a$ deviates from the macrospin expectation, Fig. 4(b)] containing 367 atoms (i.e., 5 nm diameter), the precession mode is found at a frequency (116 GHz) lower than the macrospin expectation and the PSD is lower at the edge atoms than at the center atoms, what can be ascribed to dipolar couplings.

More interestingly a second peak at larger frequency has been resolved for the larger dot [Fig. 4(b)]. This peak does not correspond to a precession mode but to a confined spin-wave mode: the precessions of two symmetric spins with respect to the dot axis show a $\pi$ phase shift and cancel when averaging the effect on the whole particle. The shape of this mode shown in the inset of Fig. 4(b) is in good agreement with a previous calculation on a similar system [17]. As spin-wave modes in a nanodot are confined, the energy needed to create inhomogeneous modes is inversely proportional to a power of the dot diameter, depending on the boundary conditions [18]. This explains why spin-wave

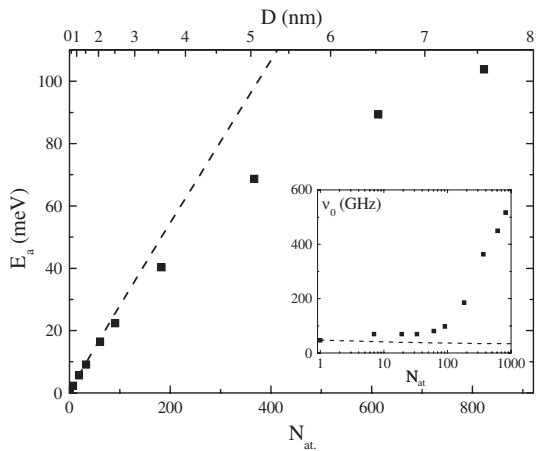

FIG. 3. Activation energy $E_a$ and (inset) attempt frequency $\nu_0$ dependence versus the nanodot size, obtained from the telegraph noise simulations (Fig. 2). The dotted lines correspond to the macrospin expectation.

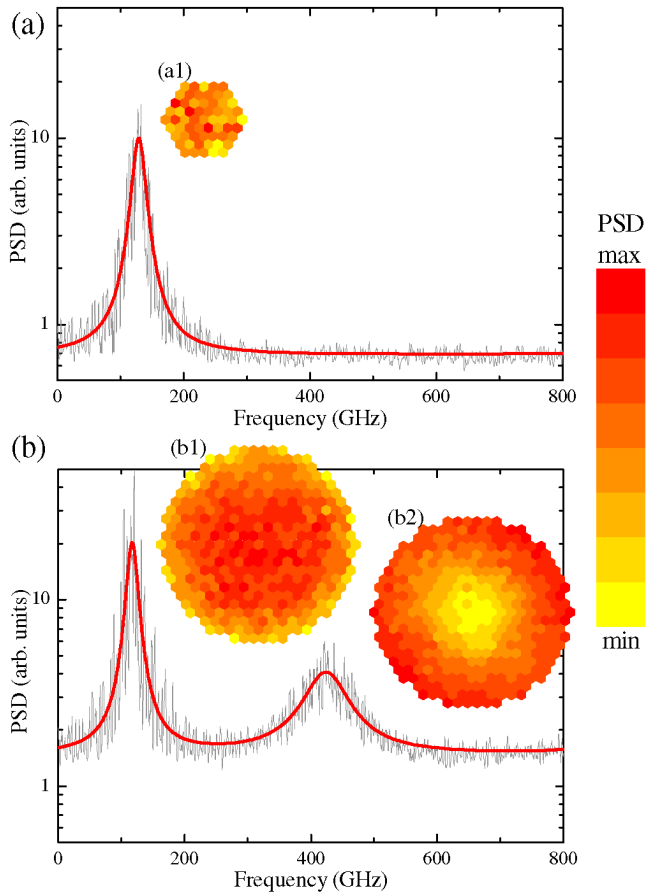

FIG. 4 (color online). Fourier transform of the in-plane component of a single spin (taken at an arbitrary position in the nanodot) extracted from the magnetization fluctuation on a (a) 61-atom nanodot at 10 K and a (b) 367-atom nanodot at 50 K. The full (red) lines are multiple Lorentzian fits. For each peak (eigenmode), the cartography of the power spectral density (PSD) is shown in inset with the color bar shown on the right. The frequencies are, respectively, (a1) 129 GHz, (b1) 116 GHz, and (b2) 423 GHz. For each cartography, the gray scale (color) amplitude is adapted in order to enhance the PSD spatial variations [(a1) min= 9, max= 11; (b1) min= 18, max= 22; (b2) min= 1.4, max= 5.5].

modes can only be observed above a critical diameter for a given temperature. As observed in the simulations (constant magnetization amplitude), the magnetization reversal is a collective effect over the whole particle so that only the precession mode is directly related to the reversal. However, other excitation modes can also play a role acting as an energy reservoir [9]. In that case, it was shown that energy exchanges between the different excitation modes can provoke an increase of the switching rate, which appears in our simulations as both an increase of $\nu_0$ and a decrease of $E_a$.

In conclusion, we have shown that the activation energy for the perpendicular magnetization thermal reversal of nanodots is not scaling linearly with the magnetic anisotropy energy, which is of considerable importance in the field of nanomagnetism and its numerous applications. Using atomic scale simulations at finite temperatures based on the Heisenberg Hamiltonian and the assumption of constant magnetic anisotropy per atom, we have pointed out the role of thermally excited confined spin-wave modes on the increase of the reversal rate as compared to the macrospin model. Further calculations could explore different parameters (for example, *ab initio* inputs of different magnetic moments and anisotropies at the border, exchange constant, shape) in order to better understand the link between the spin-wave excitations and the reversal rate. Ferromagnetic resonance on particle assemblies [19] as well as spin polarized STM experiments could allow study of the mapping of spin-wave modes in nanostructures [20] and their correlation with the thermal reversal process [5].

We thank W. Wulfhekel, J. Miltat, H. Hurdequint, and J.-C. Lévy for stimulating discussions. We acknowledge financial support from the French ministry of research (young researcher ACI), the CEFIPRA (Indo-French centre for the promotion of advanced research), the ANR (National Research Agency), the CNRS, and the Région Île-de-France (CNANO and SESAME).